\documentclass{IEEEtran}
\usepackage{amssymb}
\usepackage{amsfonts}
\usepackage{amsmath}
\usepackage{algorithm}
\usepackage{graphicx,subfigure,cite,multicol,multirow,diagbox,booktabs,array}

\usepackage[dvips]{color}
\usepackage{float}

\ifCLASSINFOpdf
\else
\fi
\begin{document}
\title{Efficient Receive Beamformers for Secure Spatial Modulation against a Malicious Full-duplex Attacker with Eavesdropping Ability}

\author{Xinyi Jiang,~Xiaoyu Liu,~Riqing Chen,~Yuntian~Wang,~Feng Shu,~and Jiangzhou Wang

\thanks{This work was supported in part by the National Natural Science Foundation of China (Nos. 61771244))(Corresponding authors: Feng Shu and Riqing Chen).}
\thanks{Xinyi Jiang,~Xiaoyu Liu,~Yuntian~Wang,~and ~Feng Shu are with School of Electronic and Optical Engineering, Nanjing University of Science and Technology, Nanjing, 210094, China.}
 \thanks{Riqing Chen and Feng Shu are also with the College of Computer and Information Sciences, Fujian Agriculture and
Forestry University, Fuzhou 350002, China (e-mail: shufeng@njust.edu.cn).}
\thanks{Feng Shu is also with the School of Information and Communication Engineering,  Hainan University,~Haikou,~570228, China.}
\thanks{Jiangzhou Wang is with the School of Engineering and Digital Arts, University of Kent, Canterbury CT2 7NT, U.K. Email: \{j.z.wang\}@kent.ac.uk.}

}
\maketitle

\begin{abstract}
In this paper, we consider a new secure spatial modulation scenario with a full-duplex (FD) malicious attacker Mallory owning eavesdropping capacity, where Mallory works on FD model and transmits a malicious jamming such as artificial noise (AN) to interfere with Bob. To suppress the malicious jamming on Bob from Mallory, a conventional maximum receive power (Max-RP) at Bob is presented firstly. Subsequently, to exploit the colored property of noise plus interference at Bob, a whitening-filter-based Max-RP (Max-WFRP) is proposed with an obvious  performance enhancement over Max-RP. To completely remove the malicious jamming from Mallory,  a Max-RP with a constraint of forcing the malicious jamming from Mallory to zero at Bob is proposed. To further improve  secrecy rate (SR) by removing the ZF contraint (ZFC),  the maximum signal-to-jamming-plus-noise ratio (Max-SJNR) is proposed. Our proposed methods have closed-form expressions. From simulation results, the four receive beamforming  methods have an increasing order in performance: Max-RP,  Max-RP with ZFC,~and Max-SJNR$\approx$Max-WFRP.  Additionally, the latter two  harvest a substantial performance gains over Max-RP and Max-RP with ZFC in the low and medium signal-to-noise ratio regions.
\end{abstract}


\IEEEpeerreviewmaketitle

\section{Introduction}
In multiple-input multiple-output (MIMO) systems, spatial modulation (SM) \cite{4382913} is a promising technology which exploits both the index of activated antenna and amplitude phase modulation (APM) symbol to transmit messages. Recently, SM has exhibited its benefits of increasing energy efficiency \cite{6678765} and avoiding the inner-channel interference (ICI). However, due to the broadcasting characteristic of wireless transmission, confidential information might be eavesdropped by illegal receivers. This is  a problem of  physical layer security \cite{7374656,7931710,7762075,8727455,7885602}. In \cite{8727455}, the author enhance the legitimate security by joint precoding optimization with and without eavesdroppong channel state information (CSI) . In \cite{7116516,8418689}, security is enhanced by emitting the artificial noise (AN) onto the null-space of the desired channel to interfere unknown eavesdropper and the latter derived the closed-form approximated expression ESR with the aid of AN in perfect and imperfect CSI, respectively. In \cite{8025351}, the authors emit the AN by a full-duplex desired receiver, where the confidential information is received and at the same time efficiently interfere the eavesdropper. Moreover, the authors in \cite{8668810} investigate the power allocation between confidential message and AN so that we can get a fixed power allocation factor to achieve higher SR performance.

As for secure SM, the authors in \cite{8373751,6733251} investigate the transmit antenna selection (TAS) schemes, the former proposed leakage-based and Max-SR schemes achieving a good balance between complexity and SR performance, and the latter selected two antennas each time to improve SR performance. In \cite{8638547}, the author proposed three active antenna-group (AAG) selection methods which performed well in the low, medium and high SNR regions with low complexity, respectively. In \cite{9064699}, the author jointly optimized the problem of maximizing SR over TAS and AN projection with joint and separate solutions. Meanwhile, the author in \cite{8768073} proposed deterministic and random antenna selection methods with the aid of zero-forcing precoding and fast radio-frequency switches to enhance the security of the PSM system.

However, in the aforementioned literature, the illegal receivers are all passive eavesdroppers, which only receive confidential message (CM) and don't emit malicious jamming to the desired receiver. In such a situation,  Bob cannot measure the CSI from Eve to Bob. It is impractical to know the CSIs from Alice to Eve and Eve to Bob in advance. If Eve becomes Mallory, which means it can send a malicious jamming towards Bob, this dilemma  will disappear. Why? Due to the fact that Mallory emits jamming,  Bob can estimate the CSI from Mallory to Bob. These estimated CSI can be used to suppress the jamming from Mallory. Similarly, Alice also obtain the CSI from Alice to Mallory, and prevent the interception from Eve with the help of AN. To simplify the derivation process,  the perfect CSIs from
 Mallory and Alice to Bob are available in the following.

Considering the worst scenario, we propose a novel secure SM (FDM-SSM) system with a FD malicious attacker Mallory having eavesdropping ability. In our FDM-SSM system,  Mallory can simultaneously send malicious jamming signal and intercept the CMs from Alice. Therefore, how to optimize the receive beamforming (RBF) such that the malicious jamming is reduced efficiently and the performance is  improved obviously is a challenging issue. This motivates us to propose several RBF methods, and our main contributions are summarized as follows:
\begin{enumerate}
 \item To suppress the malicious interference from Mallory and achieve a high  performance, the traditional maximum receive power (Max-RP) at Bob method is first presented via the system.  However, consider the noise plus interference at Bob includes three parts: AN from Alice, malicious jamming from Mallory and receiver noise at Bob, which obviously  is  colored,  the conventional Max-RP doesn't  exploit this colored property. Therefore, we propose an improved whitening-filter-based Max-RP (Max-WFRP). First,  we compute the covariance matrix of colored noise plus interference at Bob, multiply the original receive vector at Bob from left by the root of the inverse of covariance matrix to whiten the colored jamming plus noise. By simulation, we find the proposed Max-WFRP harvests  substantial performance gains over the conventional Max-RP in terms of SR and BER.
 \item To completely eliminate the malicious jamming signal from Mallory, a Max-RP with zero-forcing  constraint (ZFC) is proposed to force the malicious jamming from Mallory to zero and meanwhile maximize the receive power of CMs at Bob. However, due to the strict ZFC, the transmit space of CMs is also reduced dramatically. To  remove the ZFC, a maximum signal-to-jamming-plus-noise ratio (Max-SJNR) is proposed to strike a good balance between suppressing jamming and improving performance. From simulation results,  the proposed Max-SJNR performs much better than the proposed Max-RP with  ZFC in terms of SR and BER performance.

    \end{enumerate}

The rest of this paper is organized as follows. Section II presents the FDM-SSM system model and gives a definition for its average SR. In Section III, the RBF schemes for Max-SR are proposed and their closed-form expression is given. In Section IV, numerical simulation results are presented. Finally, we draw our conclusions in Section V.

\emph{Notations:} Boldface lower case and upper case letters denote vectors and matrices, respectively. $(\cdot)^{H}$ denotes conjugate and transpose operation.  $\mathbb{E}\{\cdot\}$ represents expectation operation. $\|\cdot\|$ denotes 2-norm.

\section{System Model}

\begin{figure}
  \centering
  \includegraphics[width=0.45\textwidth]{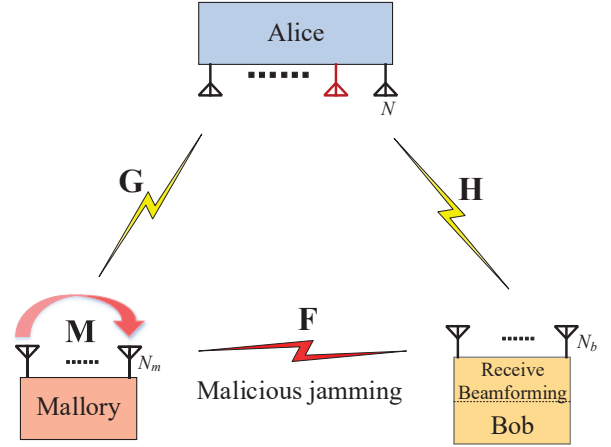}\\
  \caption{Block diagram for secure FDM-SSM network.}\label{fig1}
\end{figure}
As shown in Fig.~\ref{fig1}, the FDM-SSM system considered consists of a legal transmitter (Alice) with $N$ transmit antennas (TAs),a receiver (Bob) with $N_b$ receive antennas (RAs), and a FD malicious (Mallory) with $N_{m}$ antennas. Here, Mallory works on FD model. In other words, he intercepts the CMs from Alice and transmits  a malicious jamming towards Bob.

In general, it is noted that $N$ is not a power of 2. For a SM system, the number of active antennas are chosen from $N$, where $N_t$ is equal to ${2}^{\lfloor\textrm{log}_2N\rfloor}$. And Alice activates one of $N_t$ TAs to emit $M$-ary APM symbol and uses the index of activated antenna to convey spatial bits. As a result, $\textrm{log}_2N_tM$ bits per channel use can be transmitted.  The transmit signal with the aid of AN from Alice and the jamming signal sent from Mallory can be expressed as
\begin{align} \label{xa}
\textbf{x}_a&=\sqrt{\beta P}\textbf{e}_ns_m+\sqrt{(1-\beta)P}\textbf{P}_{\rm{AN}}\textbf{n}_{a},
\end{align}
and
\begin{align} \label{xe}
\textbf{x}_m&=\sqrt{{P}_M}\textbf{P}_{\rm{JM}}\textbf{n}_{m},
\end{align}
respectively, where $\beta \in [0,1]$ is the power allocation factor and $P$ denotes the total transmit power of Alice. Accordingly, $\textbf{e}_n$ is the $n$th column of $\textbf{I}_{N_{t}}$, indicating that the $i$th antenna is chosen to transmit symbol $s_m$, which is equiprobably drawn from discrete $M$-ary APM constellation for $m\in\left\{1,2,\cdots,M\right\}$ with $\mathbb E|s_m|^2=1$. Moreover, matrix $\textbf{P}_{\rm{AN}}\!\in\! \mathbb{C}^{N_t \times N_t}$ is the AN projection matrix and $\textbf{n}_{a}\sim\mathcal{CN}(0,\sigma_a^2\textbf{I}_{N_t})$ is the  AN vector. In (2), $P_M$ is the transmit power of Mallory and $\textbf{P}_{\rm{JM}}\!\in\! \mathbb{C}^{N_{m} \times N_{m}'}$ is the transmit beamforming matrix with $N_m'\leq N_m$ of jamming vector $\textbf{n}_{m}\sim\mathcal{CN}(0,\sigma_m^2\textbf{I}_{N_{m}})$.

Let $\textbf{H}\in \mathbb{C}^{N_b \times N}$, $\textbf{G}\in \mathbb{C}^{N_{m} \times N}$, $\textbf{F}\in \mathbb{C}^{N_b \times {N_{m}}}$ and $\textbf{M}\in \mathbb{C}^{N_{m} \times {N_{m}}}$ denote the channel gain matrices from Alice  to Bob, from Alice to Mallory,  from Mallory to Bob,  and Mallory's self-interference, respectively. Then the receive signal at Bob and Mallory after receive beamforming can be respectively expressed as
\begin{align} \label{yb}\nonumber
\textbf{y}_b=\sqrt{\beta P}\textbf{u}_{br}^{\rm{H}}\textbf{HT}\textbf{e}_ns_m+\sqrt{(1-\beta)P}\textbf{u}_{br}^{\rm{H}}\bullet \nonumber\\
\underbrace{\left(\textbf{HT}\textbf{P}_{\rm{AN}}\textbf{n}_{a}+\sqrt{P_M}\textbf{F}\textbf{P}_{\rm{JM}}\textbf{n}_{m}+\textbf{n}_b\right)}_{\textbf{w}},
\end{align}
and
\begin{align} \label{ye}\nonumber
\textbf{y}_e&=\sqrt{\beta P}\textbf{u}_{er}^{\rm{H}}\textbf{GT}\textbf{e}_ns_m+\sqrt{(1-\beta)P}\textbf{u}_{er}^{\rm{H}}\textbf{GT}\textbf{P}_{\rm{AN}}\textbf{n}_a \\ & +\sqrt{P_M}\textbf{u}_{er}^{\rm{H}}\textbf{M}\textbf{P}_{\rm{JM}}\textbf{n}_m+\textbf{u}_{er}^{\rm{H}}\textbf{n}_e,
\end{align}
where $\textbf{u}_{br}\in \mathbb{C}^{N_b \times 1}$ and $\textbf{u}_{er}\in \mathbb{C}^{N_{er} \times 1}$ are the receive beamforming vectors at Bob and Mallory. In addition, $\textbf{n}_b\sim\mathcal{CN}(0,\sigma_b^2\textbf{I}_{N_b})$ and $\textbf{n}_e\sim\mathcal{CN}(0,\sigma_e^2\textbf{I}_{N_e})$ denote the complex additive white Gaussian noise (AWGN) vectors at Bob and Mallory, respectively. Meanwhile, $\textbf{T} \in \mathbb{R}^{N\times{N_t}}$ is the transmit antennas selection matrix  constituted by the specially selected ${N_t}$ columns of $\textbf{I}_{N}$, which is determined by the leakage-based method in \cite{8373751}.  In (4), $\sqrt{P_M}\textbf{u}_{er}^{\rm{H}}\textbf{M}\textbf{P}_{\rm{JM}}\textbf{n}_m$ is the self-interference observed at Mallory. Consider Mallory design the matrix $\textbf{P}_{\rm{JM}}$ by null-space projection method, i.e., $\textbf{P}_{\rm{JM}}$ is the null-space of the RBF $\textbf{u}_{er}$ with ${N}_{m}'\leq {N}_{m}-1$. As a result, Mallory can eliminate the self-interference.
%The maximum-likelihood (ML) detector at Bob is given by
%\begin{align} \label{ML}
%[\hat{i},\hat{j}]=&\mathop{\textrm{arg} \ \textrm{min}}\limits_{n,m} \ \|\textbf{y}_b-\sqrt{\beta P}\textbf{u}_{br}^H\textbf{H}\textbf{e}_{n}s_m\|^2.
%\end{align}
The average SR is given as
\begin{align}
\overline{R_s}=\mathbb{E}_{\textbf{H},\textbf{G}}[I(\textbf{x};\textbf{y}_b)-I(\textbf{x};\textbf{y}_e)]^+,
\end{align}
where
\begin{align}  \nonumber
I(\textbf{x};\textbf{y}_{\rm{b}}')&=\textrm{log}_2N_tM - \left\{N_tM\right\}^{-1}\times \\ & \sum \limits_{i = 1}^{N_tM} {\mathbb{E}_{\textbf{n}_b'} \left\{ \textrm{log}_2 \sum \limits_{j=1} ^{N_tM} \textrm{exp} \left(-f_{b,i,j}+\|\textbf{n}_b'\|^2 \right) \right\}},
\end{align}
and
\begin{align}  \nonumber
I(\textbf{x};\textbf{y}_{\rm{e}}')&=\textrm{log}_2N_tM -\left\{N_tM\right\}^{-1}\times \\ & \sum \limits_{i = 1}^{N_tM} {\mathbb{E}_{\textbf{n}_e'} \left\{ \textrm{log}_2 \sum \limits_{j=1} ^{N_tM} \textrm{exp} \left(-f_{e,m,k}+\|\textbf{n}_e'\|^2 \right) \right\}},
\end{align}
where $\textbf{y}'_b\!\!=\!\!\textbf{W}_{\rm{B}}^{-1/2}\textbf{y}_b$. $f_{b,i,j}\!\!\!=\!\!\!\|\sqrt{\beta P}\textbf{W}_{\rm{B}}^{-1/2}\textbf{u}_{br}^H\textbf{H}\textbf{d}_{ij}\!\!+\!\!\textbf{n}'_b\|^2$, $\textbf{y}'_e\!\!=\!\!\textbf{W}_{\rm{E}}^{-1/2}\textbf{y}_e$, $f_{e,m,k}\!\!=\!\!\|\sqrt{\beta P}\textbf{W}_{\rm{E}}^{-1/2}\textbf{u}_{er}^H\textbf{G}\textbf{d}_{mk}+\textbf{n}'_e\|^2$, $\textbf{d}_{ij}\!\!=\!\!\textbf{x}_i\!-\!\textbf{x}_j$ and $\textbf{d}_{mk}\!\!=\!\!\textbf{x}_m\!-\!\textbf{x}_k$. Herein, $\textbf{x}_i$, $\textbf{x}_j$, $\textbf{x}_m$, and $\textbf{x}_k$ are the possible transmit vectors. Besides, $\textbf{n}'_b\!\!=\!\!\textbf{W}_{\rm{B}}^{-1/2}\textbf{U}_{br}^H\textbf{n}_b$, and $\textbf{n}'_e\!=\!\textbf{W}_{\rm{E}}^{-1/2}\textbf{U}_{er}^H\textbf{n}_e$.$\textbf{W}_{\rm{B}}$ and $\textbf{W}_{\rm{E}}$ are the covariance matrices of interference plus noise of Bob and Mallory respectively, where
\begin{align}
\textbf{W}_{\rm{B}}&=
(1-\beta)P\sigma_a^2\textbf{u}_{br}^{\rm{H}}\textbf{HT}\textbf{P}_{\rm{AN}}\textbf{P}_{\rm{AN}}^{\rm{H}}\textbf{T}^{\rm{H}}\textbf{H}^{\rm{H}}\textbf{u}_{br}\nonumber\\&+P_M\sigma_m^2\textbf{u}_{br}^{\rm{H}}\textbf{F}\textbf{P}_{\rm{JM}}\textbf{P}_{\rm{JM}}^{\rm{H}}\textbf{F}^{\rm{H}}\textbf{u}_{br}+\sigma_b^2,
\end{align}
and
\begin{align}
\textbf{W}_{\rm{E}}&=(1-\beta)P\sigma_a^2\textbf{u}_{er}^{\rm{H}}\textbf{GT}\textbf{P}_{\rm{AN}}\textbf{P}_{\rm{AN}}^{\rm{H}}\textbf{T}^{\rm{H}}\textbf{G}^{\rm{H}}\textbf{u}_{br}\nonumber\\&+P_M\sigma_m^2\textbf{u}_{er}^{\rm{H}}\textbf{M}\textbf{P}_{\rm{JM}}\textbf{P}_{\rm{JM}}^{\rm{H}}\textbf{M}^{\rm{H}}\textbf{u}_{er}+\sigma_e^2.
\end{align}
According to \cite{7116516}, it is known that premultiplying $\textbf{y}_{b}$ by $\textbf{W}_{\rm{B}}^{-1/2}$ will not change the mutual information, which results in $I(\textbf{x};\textbf{y}_b)\!=\!I(\textbf{x};\textbf{y}'_b)$ and $I(\textbf{x};\textbf{y}_e)\!=\!I(\textbf{x};\textbf{y}'_e)$.
%%The optimization problem can be casted as
%%\begin{align}\label{OSR}
%%&\max_{\textbf{u}_{br}} ~~  R_s(\textbf{u}_{br})\\ \nonumber
%&\textrm{subject} \ \textrm{to} \ \|\textbf{u}_{br}\|^2=1,
%\end{align}
%which
\section{Proposed Three Receive Beamforming Schemes at Bob}
For the newly proposed FDM-SSM system, the design of receive beamforming at Bob is very important to improve the system performance. In this section, the conventional Max-RP is presented, and three high-performance RBF schemes are proposed as follows: Max-WFRP, Max-RP with ZFC, and Max-SJNR. Finally, complexity comparison is made among them.
\subsection{Conventional Max-RP}
In this subsection, we derive the Max-RP from Max-SJNR rule. According to the definition of SJNR, we first write the SJNR at Bob as follows
\begin{align}\label{SNR}\nonumber
\textrm{SJNR}&=\frac {\beta P\mathbb E[\textbf{u}_{br}^H\textbf{H}\textbf{T}\textbf{e}_ns_ms_m^{*}\textbf{e}_n^H\textbf{T}^H\textbf{H}^H\textbf{u}_{br}]}{\mathbb E[\textbf{u}_{br}^H\textbf{w}\textbf{w}^H\textbf{u}_{br}]}\\&
=\frac {\beta P\textbf{u}_{br}^H\textbf{H}\textbf{T}\textbf{e}_n\mathbb E[s_ms_m^{*}]\textbf{e}_n^H\textbf{T}^H\textbf{H}^H\textbf{u}_{br}}{\textbf{u}_{br}^H\mathbb E[\textbf{w}\textbf{w}^H]\textbf{u}_{br}}
\end{align}
%where
%\begin{align}
%\textbf{w}=\sqrt{(1-\beta)P}\textbf{H}\textbf{T}\textbf{P}_{\rm{AN}}\textbf{n} +\sqrt{P_e}\textbf{F}\textbf{P}_{\rm{JM}}\textbf{m}+\textbf{n}_b
%\end{align}
As mentioned in Section II, $\mathbb E[s_ms_m^{*}]=1$, $\textbf{u}_{br}^H\textbf{u}_{br}=1$ and $\textbf{e}_n$ is the $n$th column of $\textbf{I}_{N_{t}}$. To simplify the above optimization,  $\textbf{w}$ is approximated as a white noise, i.e., $\mathbb E[\textbf{w}\textbf{w}^H]=\sigma_\omega^2\textbf{I}_{N_b}$, (\ref{SNR}) can be rewritten as
\begin{align}
\textrm{SJNR}&=\beta P\left(\sigma_\omega^2N_t\right)^{-1}\textbf{u}_{br}^H\textbf{H}\textbf{T}\textbf{T}^H\textbf{H}^H\textbf{u}_{br}
\end{align}
Hence, the optimization problem of Max-SJNR reduces to
\begin{align}\label{MRC-Opt}
\max_{\textbf{u}_{br}}~~\beta P\left(\sigma_\omega^2N_t\right)^{-1}\textbf{u}_{br}^H\textbf{H}\textbf{T}\textbf{T}^H\textbf{H}^H\textbf{u}_{br}~~\textrm{s.t.}~~\textbf{u}_{br}^H\textbf{u}_{br}=1,
\end{align}
which actually is a Max-RP. The Lagrangian function associated with the above optimization is defined as
\begin{align}\label{Lag-func}
L(\textbf{u}_{br},\lambda)=\beta P\left(\sigma_\omega^2N_t\right)^{-1}\textbf{u}_{br}^H\textbf{H}\textbf{T}\textbf{T}^H\textbf{H}^H\textbf{u}_{br}-\lambda\textbf{u}_{br}^H\textbf{u}_{br}
\end{align}
whose first-order derivative with respect to $\textbf{u}_{br}$  is set to be zero as follows
\begin{align}
\frac{\partial~L(\textbf{u}_{br},\lambda)}{\partial\textbf{u}_{br}}=\frac{2\beta P}{\sigma_\omega^2N_t}\textbf{H}\textbf{T}\textbf{T}^H\textbf{H}^H\textbf{u}_{br}-\lambda\textbf{u}_{br}=\textbf{0}
\end{align}
which means the fact that $\textbf{u}_{br}$ is the eigenvector corresponding to the largest eigen-value of the matrix $\textbf{H}\textbf{T}\textbf{T}^H\textbf{H}^H$.
\subsection{Proposed Max-WFRP}
 In fact, the malicious jamming signal plus interference plus noise $\textbf{w}$ is colored not white. Therefore, the conventional Max-RP method in Subsection A may not perform optimally in terms of SNR. In what follows, we propose the Max-WFRP method, which first whitens the colored noise plus interference at Bob using their covariance matrix, and then apply Max-RP method to maximize the average SNR.

As  $\textbf{n}_{a}$, $\textbf{n}_m$ and $\textbf{n}_b$ are the independent and identically distributed i.e.,i.i.d, random vectors,~$\textbf{w}$  has a mean vector of all-zeros and covariance matrix
\begin{align}\label{R-w}
\textbf{R}_{w}\!=&\!(1-\beta) P\sigma_a^2\textbf{HT}\textbf{P}_{\rm{AN}}\textbf{P}_{\rm{AN}}^H\textbf{T}^H\textbf{H}^H \!+\!{P_M}\sigma_m^2\textbf{F}\textbf{P}_{\rm{JM}}\textbf{P}_{\rm{JM}}^H\textbf{F}^H\!\nonumber\\~~~~~~~~+&\!\sigma_b^2\textbf{I}_{N_b}.
\end{align}
Observing (\ref{R-w}), it is evident that the covariance matrix $\textbf{R}_b$ is a positive definite matrix, its eigenvalue decomposition (EVD) is in the form
%\begin{align}
$\textbf{R}_{w}=\textbf{U}_w\Lambda\textbf{U}_w^H$,
%\end{align}
where $\textbf{U}_w$ is an $N_b\times N_b$ unitary  matrix, and $\Lambda=$ is a diagonal matrix $\text{diag}({d_1},\cdots {d}_{N_b})$  with ${d}_i$ being the $i$th eigenvalue of matrix $\textbf{R}_{w}$. Let us define the WF matrix satisfying $\textbf{W}_{\rm{WF}}\textbf{W}_{\rm{WF}}^{H}=\textbf{R}_{w}^{-1}$, which yields
\begin{align}\label{WF-matrix}
\textbf{W}_{\rm{WF}}=(\textbf{U}_w\Lambda^{\frac{1}{2}})^{-1}=\Lambda^{-\frac{1}{2}}\textbf{U}_w^H,
\end{align}
where $\Lambda^{\frac{1}{2}}=\text{diag}\left(\sqrt{d_1},\cdots \sqrt{{d}_{N_b}}\right)$. Applying the WF matrix in (\ref{WF-matrix})  on $\textbf{y}_{B}=\sqrt{\beta P}\textbf{H}\textbf{T}\textbf{e}_ns_m+\omega$ yields a new system model,
\begin{align}\nonumber
\textbf{y}_{\rm{B}}'&=\sqrt{\beta P}\textbf{W}_{\rm{WF}}\textbf{H}\textbf{T}\textbf{e}_ns_m+\textbf{W}_{\rm{WF}}\textbf{w}\nonumber\\&=\sqrt{\beta P}\textbf{W}_{\rm{WF}}\textbf{H}\textbf{T}\textbf{e}_ns_m+\textbf{w}'
\end{align}
where $\textbf{w}'$ has covariance matrix $\mathbb E(\textbf{w}'\textbf{w}'^H)=\textbf{I}_{N_b}$. Similar to (\ref{MRC-Opt}), we have the form of  Max-WFRP
\begin{align}
\max_{\textbf{u}_{br}}~\beta P{N_t}^{-1}\textbf{u}_{br}^H\textbf{W}_{WF}\textbf{H}\textbf{T}\textbf{T}^H\textbf{H}^H\textbf{W}_{WF}^H\textbf{u}_{br}
~~\textrm{s.t.}~~\textbf{u}_{br}^H\textbf{u}_{br}=1.
\end{align}
Hence, $\textbf{u}_{br}$ can be set to be the eigenvector to the largest eigen-value of the matrix  $\textbf{W}_{F}\textbf{H}\textbf{T}\textbf{T}^H\textbf{H}^H\textbf{W}_{F}^H$.
\subsection{Proposed Max-RP with ZFC}
In order to completely eliminate the malicious jamming from Mallory, in that follows, we propose the ZF method of maximizing the receive power of confidential messages at Bob by projecting  the malicious jamming  signal onto the null-space of channel matrix from Alice to Bob.  The corresponding  Max-RP with ZFC optimization problem  is formalized as
\begin{align}\label{Max-RP-ZF}
&\max_{\textbf{u}_{br}} ~~  \frac {\beta P}{N_t}\textbf{u}_{br}^H\textbf{HT}\textbf{T}^H\textbf{H}^H\textbf{u}_{br} \\ \nonumber
&\textrm{s.t.}~~~\textbf{u}_{br}^H\textbf{F}\textbf{P}_{\rm{JM}}=\textbf{0}_{1\times N_m'}~~(\textrm{ZFC}),~\textbf{u}_{br}^H\textbf{u}_{br}=1.
\end{align}
To simplify the above optimization, let us define a new matrix $\textbf{F}'=\textbf{F}\textbf{P}_{\rm{JM}}$, and its singular value decomposition (SVD) is
\begin{align}
\textbf{F}'=[\mathbf{U}_s\ \mathbf{U}_\perp][\Lambda\ \textbf{0}]^H\mathbf{V},
\end{align}
where $\mathbf{U}_\perp$ spans the null-space of the column space of matrix $\textbf{F}'$. Now, to remove the first constraint,  let  us introduce a new beamforming vector $\textbf{u}_{br}'$ as follows
%\begin{align}
$\textbf{u}_{br}=\mathbf{U}_\perp\textbf{u}_{br}'$,
%\end{align}
which satisfies $\textbf{u}_{br}^H\textbf{F}'=\textbf{0}_{1 \times N_{m}'}$. Substituting the above equation in $(\ref{Max-RP-ZF})$  forms the following unstrained optimization
\begin{align}
&\max_{\textbf{u}_{br}} ~~\frac{\textbf{u}_{br}'^H\beta P\mathbf{U}_\perp^H\textbf{HT}\textbf{T}^H\textbf{H}^H\mathbf{U}_\perp\textbf{u}_{br}'}{\textbf{u}_{br}'^H\mathbf{U}_\perp^H\mathbf{U}_\perp\textbf{u}_{br}'}.
\end{align}
Using the generalized Rayleigh-Ritz theorem, the new beamforming vector $\textbf{u}_{br}'^H$  is directly equal to the eigenvector corresponding to the largest eigen-value of the matrix
\begin{align}
\textbf{A}=(\mathbf{U}_\perp^H\mathbf{U}_\perp)^{-1}(\beta P\mathbf{U}_\perp^H\textbf{HT}\textbf{T}^H\textbf{H}^H\mathbf{U}_\perp),
\end{align}
i.e., $\boldsymbol{\eta}(\textbf{A})$. Accordingly, the optimal value of $\textbf{u}_{br}$ is given by
\begin{align}
\textbf{u}_{br}=\mathbf{U}_\perp\boldsymbol{\eta}(\textbf{A}).
\end{align}
\subsection{Proposed Max-SJNR}
The ZF constraint in (\ref{Max-RP-ZF}) is extremely strict and  reduces the degrees of freedom of transmit space of CMs. Below, after this constraint is removed, a Max-SJNR is proposed to maximize the average SJNR,
\begin{align}
\max_{\textbf{u}_{br}} ~~  \textrm{SJNR}(\textbf{u}_{br})~~~\textrm{s.t.}~~~\textbf{u}_{br}^H\textbf{u}_{br}=1
\end{align}
which can make a good balance between reducing the malicious jamming and improving the performance, where $\textrm{SJNR}$ is defined as
\begin{align}
\!\!\!\frac{\frac{1}{N_t}\beta P\textbf{u}_{br}^H\textbf{H}\textbf{T}\textbf{T}^H\textbf{H}^H\textbf{u}_{br}}{ \textbf{u}_{br}^H\!(\!{P}_{M}\sigma_m^2\!\textbf{F}\textbf{P}_{\rm{JM}}\!\textbf{P}_{\rm{JM}}^H\!\textbf{F}^H\!\!\!\!+\!\!(\!1\!\!-\!\!\beta\!)\!P\!\sigma_a^2\!\textbf{H}\textbf{T}\textbf{P}_{\rm{AN}}\!\textbf{P}_{\rm{AN}}^H\!\textbf{T}^H\textbf{H}^H\!\!\!\!\!\!+\!\!\sigma_b^2\!\textbf{I}_{N_b}\!)\!\textbf{u}_{br}},
\end{align}
Using the generalized Rayleigh-Ritz theorem, the beamforming vector $\textbf{u}_{br}$ of maximizing the SJNR can be directly obtained from the eigenvector corresponding to the largest eigen-value of the matrix
\begin{align}\nonumber
&[{P}_{M}\sigma_m^2\textbf{F}\textbf{P}_{\rm{JM}}\textbf{P}_{\rm{JM}}^H\textbf{F}^H\!+\!(\!1\!-\!\beta\!)P\sigma_a^2\textbf{HT}\textbf{P}_{\rm{AN}}\textbf{P}_{\rm{AN}}^H\textbf{T}^H\textbf{H}^H\!\\&~~~~~~~~+\!\sigma_b^2\textbf{I}_{N_b}]^{-1}\times(\beta P\textbf{H}\textbf{T}\textbf{T}^H\textbf{H}^H).
\end{align}
Actually, the Max-SJNR is equivalent to the Max-WFRP proposed by us, which will be verified in the next section.
\subsection{Computational Complexity Analysis}
%In this subsection, the approximate computational complexities of the four RBF methods are analyzed. First, the complexity of the traditional MRC method is about $\mathcal{C}_{\rm{Max-RP}}=\mathcal{O}(129{N}_{b}^3)$ floating-point operations (FLOPs). For the proposed WFMRC, its computational complexity is approximated as  $\mathcal{C}_{\rm{Max-WFRP}}=\mathcal{O}(266{N}_{b}^3+3{N}_{b})$. Accordingly, the complexity of Max-RP with ZFC method is $\mathcal{C}_{\rm{ZF}}=\mathcal{O}(259{N}_{b}^3)$. Besides, the computational complexity of the proposed Max-SJNR is approximated expressed as$ \mathcal{C}_{\rm{Max-SJNR}}=\mathcal{O}(268{N}_{b}^3+3{N}_{b})$, which is slightly higher than proposed WFMRC method. Generally, their complexities have an increasing order as follows: MRC, Max-RP with ZFC, WFMRC, Max-SJNR.
Now, the approximate computational complexities of the four RBF methods are analyzed. First, the complexity of the traditional Max-RP method is about $\mathcal{C}_{\rm{Max-RP}}=\mathcal{O}(129{N}_{b}^3)$ floating-point operations (FLOPs). For the proposed Max-WFRP, its computational complexity is approximated as  $\mathcal{C}_{\rm{Max-WFRP}}=\mathcal{O}(266{N}_{b}^3+3{N}_{b})$. Accordingly, the complexity of Max-RP with ZFC method is $\mathcal{C}_{\rm{Max-RP with ZFC}}=\mathcal{O}(259{N}_{b}^3)$. Besides, the computational complexity of the proposed Max-SJNR is approximated  as$ \mathcal{C}_{\rm{Max-SJNR}}=\mathcal{O}(268{N}_{b}^3+3{N}_{b})$, which is slightly higher than proposed Max-WFRP method. Generally, their complexities have an increasing order as follows: Max-RP, Max-RP with ZFC, Max-WFRP, Max-SJNR.

\section{Simulation and Discussion}
In this section, numerical simulations are presented to make a performance comparison among three proposed methods and conventional Max-RP from three different aspects: average SR, cumulative density function (CDF) of SR and the BER. Simulation parameters are set as follows: $N_b=6$, $N_t=8$, $P=10$W, $\sigma_b^2=\sigma_e^2$, and  modulation scheme being QPSK.
\begin{figure}[h]
\centerline{\includegraphics[width=0.47\textwidth]{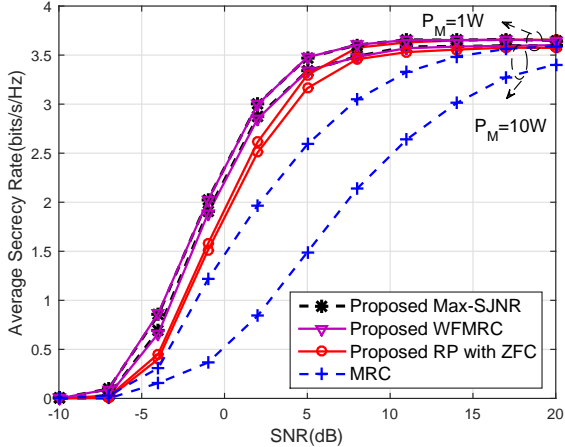}}
\caption{Curves of average SR versus SNR for different $P_M$ of Mallory.}
\label{fig2}
\end{figure}

Fig.~\ref{fig2} plots the curves of SR versus SNR for two different interference power: $P_M=1$W and $P_M=10$W. From Fig.~\ref{fig2}, it is seen that  the proposed Max-WFRP and Max-SJNR methods perform much better than Max-RP with ZFC and Max-RP. The conventional Max-RP is the worst one among all beamformers in terms of SR due to the fact that it omits the colored property of noise plus interference at Bob. Conversely, Max-WFRP achieves the best one among four methods.  Additionally, the SR performance of the proposed Max-RP with ZFC method approaches those of Max-SJNR, Max-WFRP, and conventional Max-RP in the high SNR region. In summary, they have an increasing order in SR performance: Max-RP, Max-RP with ZFC, and Max-WFRP$\approx$ Max-SJNR.
\begin{figure}[h]
\centerline{\includegraphics[width=0.47\textwidth]{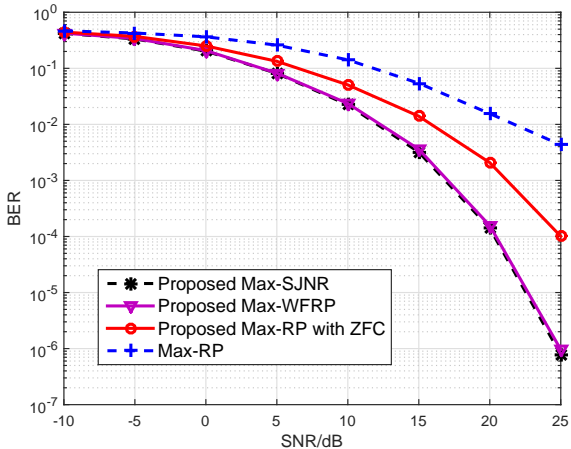}}
\caption{Average BER curves of Bob for four RBF methods with $\textrm{P}_{M}=1$W.}
\label{fig3}
\end{figure}
\begin{figure}[h]
\centerline{\includegraphics[width=0.47\textwidth]{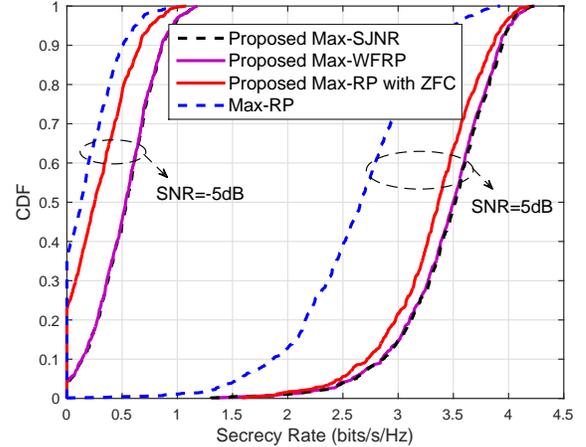}}
\caption{CDF curves of SR for SNR=-5dB, 5dB and $\textrm{P}_{M}=1$W.}
\label{fig4}
\end{figure}

Fig.~\ref{fig3} illustrates the curves of average BER versus SNR for the above four methods. From Fig.~\ref{fig3}, it is seen that the proposed Max-SJNR and the proposed Max-WFRP  still obviously exceed the remaining two methods Max-RP with ZFC and conventional Max-RP in terms of BER performance. Similarly, Max-SJNR is still overlapped with Max-WFRP and the performance difference between them is trivial. In particular, as SNR increases, the BER performance gains  achieved by Max-WFRP and Max-SJNR over Max-RP with ZFC and conventional Max-RP become more significant. Similar to Fig.~\ref{fig2}, there is the same increasing order in BER performance: Max-RP, Max-RP with ZFC, and Max-WFRP$\approx$ Max-SJNR. Fig.~\ref{fig4} shows the CDF curves of SR for the four  methods with two different value of SNRs: -5dB and 5dB. From Fig.~\ref{fig4} , we find the same performance tendency as shown in Fig.~\ref{fig3}.
%This is due to the fact the MRC method doesn't eliminate the jamming signals from Mallory, and in the low SNR regions, the BER performance is affected by the noise mainly, however, in the high SNR regions, the impact of jamming is more crucial to the BER performance. Hence, it is effective to whiten the noise before MRC.
\section{Conclusion}
In this paper, we have made an investigation of RBF methods in the newly proposed SSM system with a malicious full-duplex attacker having an eavedropping ability. First, the conventional Max-RP method was  derived to design RBF, after that, three high-performance RBF design methods, Max-WFRP, Max-RP with  ZF constraint, and Max-SJNR have been proposed to improve the SR performance of the system. From simulation results, the proposed Max-WFRP  achieves the same performance as Max-SJNR. They are much better than conventional Max-RP and Max-RP with  ZFC  in terms of SR and BER performance and have the same order computational complexity as Max-RP and Max-RP with ZFC. Interestingly, the proposed Max-WFRP has a slightly lower complexity compared to the proposed Max-SJNR.
 %we have made an investigation on receive beamforming design methods for the FDM-SSM systems. Here we propose four RBF design methods....
%\begin{appendix}
%\subsection{Proof of Covariance in (14)}
%Since the colored noise written in (13) has zero mean, the covariance can be written as
%\begin{align}\nonumber
%\textbf{cov}[\textbf{n}_{B},\textbf{n}_{B}]&=\mathbb E[\textbf{n}_{B}\textbf{n}_{B}^H]\\&\nonumber=\mathbb E[(\sqrt{(1-\beta)P}\textbf{H}\textbf{P}_{\rm{AN}}\textbf{n} +\sqrt{P_e}\textbf{F}\textbf{P}_{\rm{JM}}\textbf{m}+\textbf{n}_b)\\&\cdot(\sqrt{(1-\beta)P}\textbf{n}^H\textbf{P}_{\rm{AN}}^H\textbf{H}^H +\sqrt{P_e}\textbf{m}^H\textbf{P}_{\rm{JM}}^H\textbf{F}^H+\textbf{n}_b)^H]
%\end{align}
%because $\textbf{n}_{b}$, $\textbf{m}$ and $\textbf{n}$ are independent and identically distributed i.e.,i.i.d, $\mathbb E[\textbf{n}_{b}\textbf{n}_{b}^H]=\sigma_b^2$, $\mathbb E[\textbf{m}\textbf{m}^H]=\sigma_n^2$, $\mathbb E[\textbf{n}\textbf{n}^H]=\sigma_n^2$, $\mathbb E[\textbf{n}_b\textbf{n}^H]=0$, $\mathbb E[\textbf{n}_b\textbf{m}^H]=0$ and $\mathbb E[\textbf{n}\textbf{m}^H]=0$. (28) can be written as
%\begin{align}\nonumber
%\textbf{cov}[\textbf{n}_{B},\textbf{n}_{B}]\!&=\!(1-\beta) P\sigma_n^2\textbf{H}\textbf{P}_{\rm{AN}}\textbf{P}_{\rm{AN}}^H\textbf{H}^H \!+\!{P_e}\sigma_m^2\textbf{F}\textbf{P}_{\rm{JM}}\textbf{P}_{\rm{JM}}^H\textbf{F}^H\!\\&+\!\sigma_b^2\textbf{I}_{N_b}
%\end{align}
%and we can obtain covariance matrix in (14).
%\end{appendix}
\ifCLASSOPTIONcaptionsoff
  \newpage
\fi
%\textbf{cov}\[\textbf{n}_{B},\textbf{n}_{B}\]=
\bibliographystyle{IEEEtran}
\bibliography{IEEEfull,refx}
%\begin{appendix}
%\end{appendix}

%\begin{align}\nonumber
%\textbf{R}_{b}=\mathbb E[\textbf{n}_{B}\textbf{n}_{B}^H]&=(1-\beta) P\sigma_n^2\textbf{H}_{k}\textbf{P}_{\rm{AN}}\textbf{P}_{\rm{AN}}^H\textbf{H}_{k}^H + P_e\sigma_m^2\textbf{F}\textbf{P}_{\rm{JM}}\textbf{P}_{\rm{JM}}^H\textbf{F}^H\\& \nonumber
%+\sigma_b^2\textbf{I}_{N_b}
%\end{align}
%\begin{align}\nonumber
%\textbf{R}_{b}^{-1}\textbf{H}_{k}\textbf{H}_{k}^H
%\end{align}
%\begin{align}\nonumber
%\textbf{W}_{F}\textbf{H}_{k}\textbf{H}_{k}^H\textbf{W}_{F}^H
%\end{align}
\end{document}